\pdfoutput=1
\documentclass[prl,twocolumn,showpacs,amsmath,amssymb,floatfix,superscriptaddress]{revtex4-1}

\usepackage{graphicx}
\usepackage{dcolumn}
\usepackage{bm}
\usepackage{hyperref}

\begin{document}

\title{Quantum spin-transfer torque induced nonclassical magnetization dynamics and electron-magnetization entanglement}

\author{Priyanka Mondal}
\affiliation{Department of Physics and Astronomy, University of Delaware, Newark, DE 19716, USA}
\author{Marko D. Petrovi\'{c}}
\affiliation{Department of Mathematical Sciences, University of Delaware, Newark,  DE 19716, USA}
\author{Petr Plech\'a\v{c}}
\affiliation{Department of Mathematical Sciences, University of Delaware, Newark,  DE 19716, USA}
\author{Branislav K. Nikoli\'{c}}
\email{bnikolic@udel.edu}
\affiliation{Department of Physics and Astronomy, University of Delaware, Newark, DE 19716, USA}

\begin{abstract} 
The standard  spin-transfer torque (STT)---where spin-polarized current drives dynamics of magnetization viewed as a classical vector---requires noncollinearity between electron spins carried by the  current and magnetization of a ferromagnetic layer. However, recent experiments [A.~Zholud \emph{et al.}, Phys. Rev. Lett. {\bf 119},  257201 (2017)] observing magnetization dynamics in spin valves at cryogenic temperatures, even when electron spin is collinear to magnetization, point at overlooked quantum effects in STT which can lead to {\em highly nonclassical}  magnetization states. Using fully quantum many-body treatment, where an electron injected as spin-polarized wave packet interacts with local spins comprising the anisotropic quantum Heisenberg ferromagnetic chain, we define {\em quantum STT as any time evolution of local spins due to initial many-body state not being an eigenstate of electron+local-spins system}. For time evolution caused by injected spin-$\downarrow$ electron scattering off local $\uparrow$-spins, {\em entanglement} between electron subsystem and local spins subsystem takes place leading to {\em  decoherence} and, therefore, shrinking of the total magnetization but without rotation from its initial orientation which explains the experiments. Furthermore, the same processes---entanglement and thereby induced decoherence---are present also in standard noncollinear geometry, together with the usual magnetization rotation. This is because STT in quantum many-body picture is caused only by electron spin-$\downarrow$ factor state, and the only difference between collinear and noncollinear geometries is in relative size of the contribution of the initial separable state containing such factor state to superpositions of separable many-body quantum states generated during time evolution.
\end{abstract}

\maketitle

The standard spin-transfer torque (STT)~\cite{Ralph2008}, predicted in the seminal work of  Slonczewski~\cite{Slonczewski1996} and Berger~\cite{Berger1996}, is a phenomenon where a flux of spin-polarized electrons injected into a ferromagnetic metal (FM) layer drives its magnetization dynamics. The origin of STT is transfer of spin angular momentum from electrons to local magnetic moments of the FM layer, so it is fundamentally a nonequilibrium quantum many-body physics effect. Nevertheless, local magnetic moments are typically treated as {\em classical vectors  of fixed length}~\cite{Ralph2008,Berkov2008} whose dynamics is governed by the Landau-Lifshitz-Gilbert (LLG) equation~\cite{Wieser2015} extended by 
adding the STT term~\cite{Manchon2008a,Petrovic2018,Ellis2017} 
\begin{equation}\label{eq:stt}
\mathbf{T} \propto \langle \hat{\mathbf{s}}_e \rangle \times \mathbf{S}(\mathbf{r}).
\end{equation}
Thus,  the nonequilibrium spin density $\langle \hat{\mathbf{s}}_e\rangle$ caused by flowing electrons {\em must be noncollinear} to the direction of local spin $\mathbf{S}(\mathbf{r})$ [i.e.,  to the local magnetization proportional to local spin], to drive magnetization dynamics in such a classical picture. The dynamics can include oscillations or complete reversal, whose conversion into  resistance variations has emerged as a key resource for next generation spintronic technologies, such as nonvolatile magnetic random access memories, microwave oscillators, microwave detectors, spin-wave emitters, memristors and artificial neural networks~\cite{Locatelli2014,Kent2015,Borders2017}. 

For example, passing current through a spin valve trilayer fixed-FM/normal-metal/free-FM, as employed in early experiments on standard STT~\cite{Tsoi2000,Katine2000}, causes first FM layer with fixed magnetization to spin-polarize the current which then impinges onto the second FM layer with free magnetization that fluctuates in the classical picture due to a random magnetic field caused by thermal motion. When impinging spins and fluctuating magnetization become noncollinear, standard STT can either amplify such fluctuations (for fixed-to-free spin current direction) or reduce them (for free-to-fixed spin current direction), as confirmed theoretically~\cite{Li2004b} and experimentally~\cite{Koch2004} at room temperature.  

However, this well-established picture {\em cannot} explain very recent experiments~\cite{Zholud2017} on collinear spin valves at cryogenic temperatures $\lesssim 3$ K, where resistance measurements have revealed magnetization dynamics even though thermal fluctuations that could introduce noncollinearity between the free and fixed magnetizations are suppressed. This implies a mechanism where standard STT is zero, $\mathbf{T} \equiv 0$ in Eq.~\eqref{eq:stt}, so that magnetization does not rotate from the the initial configuration. Nevertheless, it changes its length, thereby signaling generation of {\em highly nonclassical} magnetization states~\cite{Zholud2017}. However, the proposed intuitive picture~\cite{Zholud2017} where such mechanism would amplify  quantum  spin fluctuations, for both fixed-to-free and free-to-fixed spin current directions, cannot be rigorously justified. That is, although quantum fluctuations of the local spin operators~\cite{Takahashi2013} (or, equivalently, zero-point energy of magnons as bosonic particles to which spin operators can be mapped) play an important role in lowering the energy of  classical ground states of antiferromagnets~\cite{Singh1990} or noncollinear spin textures~\cite{Roldan-Molina2015}, they {\em vanish} in a FM with uniaxial anisotropy because the collinear state of local magnetic moments is also a ground eigenstate of the exact Hamiltonian~\cite{Yosida1996}.  
 
Aside from few disparate attempts~\cite{Wang2012f,Gauyacq2011,Mahfouzi2017}, a general framework for describing current-driven {\em quantum dynamics} of magnetization is lacking.
Note that quantum transport theories, such as the nonequilibrium Green function formalism~\cite{Haney2007,Petrovic2018,Nikolic2018,Ellis2017} or the scattering matrix approach~\cite{Stiles2002,Wang2008b}, are routinely used to compute $\langle \hat{\mathbf{s}}_e\rangle$ in Eq.~\eqref{eq:stt} for  a given single-particle Hamiltonian, but this serves only as an input~\cite{Petrovic2018,Ellis2017} for the LLG equation describing classical dynamics of magnetization. The LLG equation can be justified under the assumptions~\cite{Wieser2015} of large spin $S \rightarrow \infty$, $\hbar \rightarrow 0$ (while $S \times \hbar  \rightarrow 1$) and in the {\em absence of entanglement}. The latter assumption means that local spins comprising the total magnetization should remain in a separable quantum state, $|S_1 \rangle \otimes |S_2 \rangle \otimes \cdots \otimes |S_N\rangle$, as exemplified by the ground state of FM, $|\!\! \uparrow \rangle \otimes |\!\! \uparrow \rangle \otimes \cdots \otimes |\!\! \uparrow\rangle$. 

\begin{figure}
	\includegraphics[scale=1.0,angle=0]{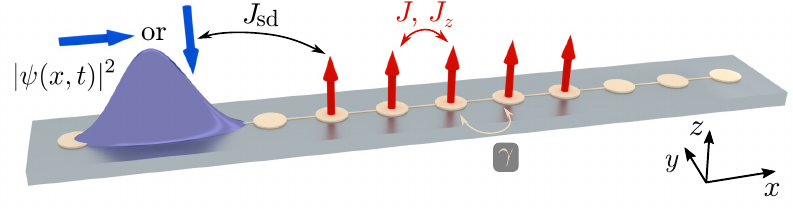}
	\caption{Schematic view of a quantum many-body system, exhibiting quantum STT, which consists of a FM layer whose local spins comprise the XXZ quantum Heisenberg ferromagnetic chain with anisotropic exchange interactions $J$ and $J_z$, and are attached to 1D TB chain where electron hops with parameter $\gamma$. The spin-polarized electron wave packet is injected along the TB chain, with its spin pointing in the $-z$- or  $+x$-direction which is collinear (a case where standard STT of Slonczewski~\cite{Slonczewski1996} and Berger~\cite{Berger1996} is absent) or noncollinear, respectively, to local spins pointing initially along the $+z$-direction.  The spin of the wave packet interacts with local spins via $s$-$d$ exchange interaction.}
	\label{fig:fig1}
\end{figure}

\begin{figure}
	\includegraphics[scale=0.52,angle=0]{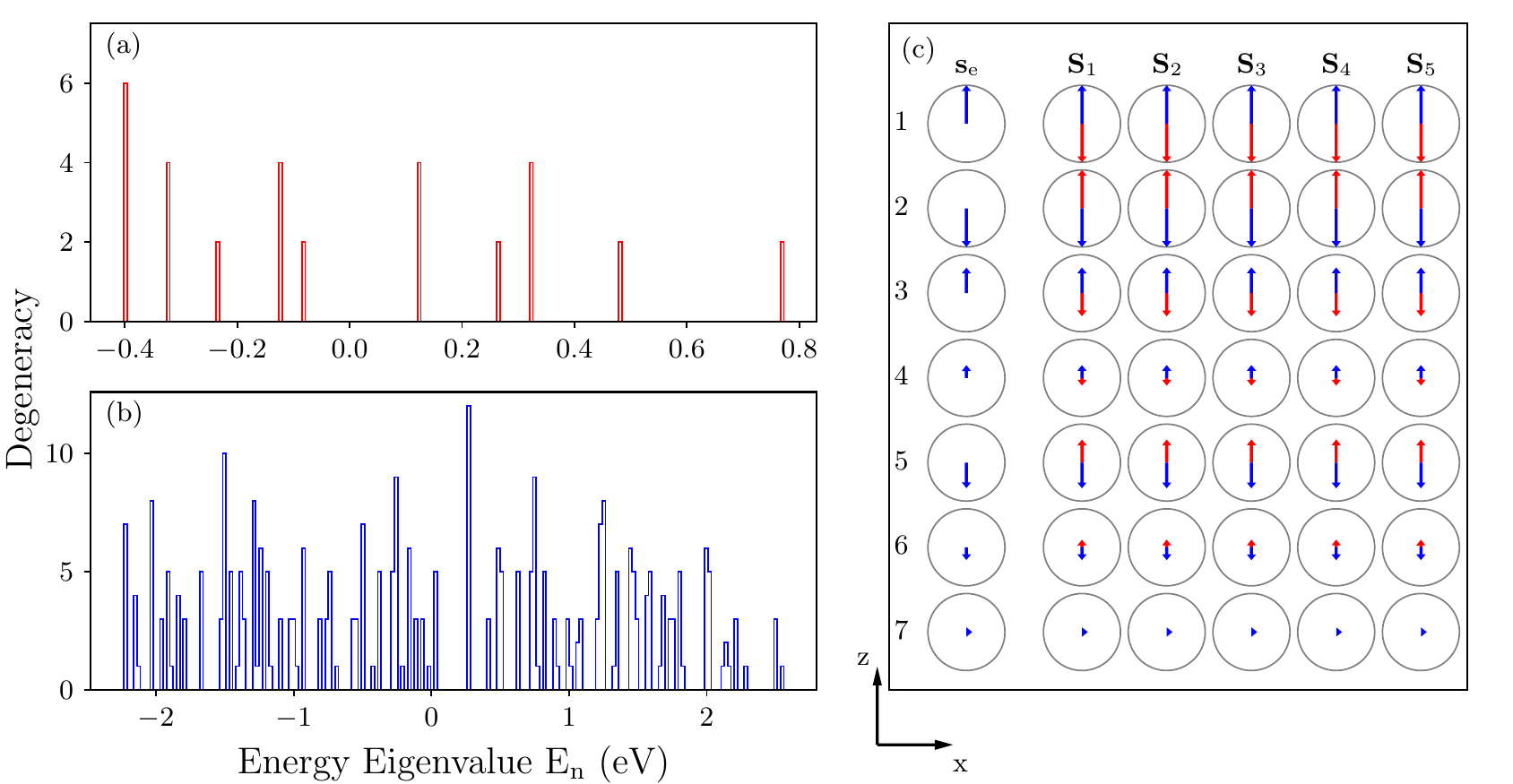}
	\caption{(a) Eigenspectrum of the XXZ quantum Heisenberg ferromagnetic chain whose $N=5$ local spins in Fig.~\ref{fig:fig1} do not interact with electron spin ($J_{sd} =0$). (b) Eigenspectrum of many-body Hamiltonian in Eq.~\eqref{eq:hamil} whose local spins interact via \mbox{$s$-$d$} interaction ($J_{sd}=0.1$ eV)  with electron spin on TB site $i$. (c) Expectation value of electron spin (first column) and local spins, extracted from their subsystem density matrices via Eq.~\eqref{eq:rhospin}, in the degenerate ground state  
		of the lowest energy in panel (a) [red arrows] or (b) [blue arrows].}
	\label{fig:fig2}
\end{figure}

Instead of classical micromagnetics~\cite{Berkov2008,Li2004b} or quantum-classical~\cite{Petrovic2018,Ellis2017} description of standard-STT-induced magnetization dynamics, here 
we introduce a quantum many-body picture of {\em both} flowing-electron-spin--local-spins interactions and the ensuing time evolution  of local spins at zero temperature. For this purpose, we employ a system depicted in Fig.~\ref{fig:fig1} where spin-polarized electron wave packet, assumed to originate from a fixed FM layer, is injected along one-dimensional (1D) tight-binding (TB) chain whose sites in the middle host local spins comprising a quantum Heisenberg ferromagnetic chain modeling the free FM layer. The states of such composite quantum system  electron+local-spins reside in the Hilbert space 
\begin{equation}\label{eq:hilbert}
\mathcal{H} = \mathcal{H}^e_\mathrm{orb} \otimes \mathcal{H}^e_\mathrm{spin} \otimes \mathcal{H}_\mathrm{spin}^1 \cdots \otimes \mathcal{H}_\mathrm{spin}^N, 
\end{equation} 
which is the tensor product of  orbital electron subspace $\mathcal{H}^e_\mathrm{orb}$ (of finite dimension equal to the number of sites of the TB chain);  two-dimensional subspace $\mathcal{H}^e_\mathrm{spin}$ for electron spin; and $\mathcal{H}_\mathrm{spin}^n$ as two-dimensional subspaces for $n=1, \cdots, N$ local spins assumed to be spin-$\frac{1}{2}$ as well.  
The system Hamiltonian  acting in $\mathcal{H}$ is
\begin{eqnarray}\label{eq:hamil}
\hat{H} & = &  - \gamma \sum_{\langle ij \rangle}  | i\rangle \langle j|  -  J_{sd} \sum_i  | i\rangle \langle i| \otimes  \hat{\mathbf{s}}_e \cdot \hat{\mathbf{S}}_i(t) \nonumber \\ 
&& -  \sum_{\langle ij \rangle} \left[J(\hat{S}_i^x \cdot \hat{S}_j^x + \hat{S}_i^y \cdot \hat{S}_j^y) + J_z \hat{S}_i^z \cdot \hat{S}_j^z \right],
\end{eqnarray}
where $| i \rangle$ is electron orbital centered on site $i$; \mbox{$\gamma=1$ eV} is hopping between nearest-neighbor sites; and \mbox{$J_{sd}=0.1$ eV} is the strength of \mbox{$s$-$d$} exchange interaction between electron and  local spins. The exchange interaction between the nearest neighbor local spins is \mbox{$J=0.1$ eV} and \mbox{$J_z=0.1005$ eV}, which are slightly different in order to include the uniaxial anisotropy. The third term in Eq.~\eqref{eq:hamil} is standardly denoted as the XXZ quantum Heisenberg ferromagnetic chain~\cite{Parkinson2010,Joel2013}. The spin operators in Eq.~\eqref{eq:hamil} are constructed as $\hat{\mathbf{s}}_e = \hat{I} \otimes \hat{\bm \sigma} \otimes \hat{I} \otimes \cdots \otimes \hat{I}$ for electron spin; $\hat{\mathbf{S}}_1 = \hat{I} \otimes \hat{I} \otimes \hat{\bm \sigma} \otimes \hat{I} \otimes \cdots \otimes \hat{I}$ for first local spins and analogously for all other local spins, where $\hat{\bm \sigma} = (\hat{\sigma}_x, \hat{\sigma}_y, \hat{\sigma}_z)$ is the vector of the Pauli matrices and  $\hat{I}$ is the unit operator. The eigenspectrum of an isolated XXZ chain (i.e., of the third term in Eq.~\eqref{eq:hamil} alone) is shown in Fig.~\ref{fig:fig2}(a), while the eigenspectrum of the whole many-body Hamiltonian in Eq.~\eqref{eq:hamil} is shown in Fig.~\ref{fig:fig2}(b). The ground state in the former (latter) case has degeneracy six (seven), as expected for coupled system of five (six) spin-$\frac{1}{2}$~\cite{Parkinson2010}. 

At $t=0$, the many-body quantum state is a separable one 
\begin{equation} \label{eq:packet}
\langle x|\Psi(t=0) \rangle  =    C  e^{ik_x x- \delta k_x^2 x^2/4} \otimes \chi_e \otimes \chi_1 \otimes \cdots \otimes \chi_N.
\end{equation}
Its first factor is electron orbital quantum state in $\mathcal{H}^e_\mathrm{orb}$ chosen as a Gaussian wave packet with momentum along the $+x$-direction and centered on the left edge of TB chain, as illustrated in Fig.~\ref{fig:fig1}, where $C$ is the normalization constant. To mimic current of electrons at the Fermi level which interact with the ground state of free FM layer within a spin valve, we use  $k_xa=0.1$ and  $\delta k_xa=0.2$ ($a$ is the lattice spacing) which specify average energy \mbox{$E=-2.36$ eV} and its standard deviation \mbox{$\delta E=0.054$ eV} for the wave packet to be close to the ground state eigenenergy \mbox{$E_0=-2.43$ eV} [Fig.~\ref{fig:fig2}(b)] of the Hamiltonian in Eq.~\eqref{eq:hamil}. In the ground state, all local spins are aligned with the anisotropy $z$-axis, as shown in Fig.~\ref{fig:fig2}(c), so we choose $\chi_n  = \left(\begin{array}{c} 1  \\ 0 \end{array} \right)$ for $n=1,\ldots,N$. To mimic minority electrons in spin valve with collinear magnetizations impinging on the free FM layer, we select initial spin polarization of the wave packet in the $-z$-direction, as described by the spinor $\chi_e  = \left(\begin{array}{c} 0  \\ 1 \end{array} \right)$. For standard STT setup with noncollinear magnetizations of the FM layers, we use spin polarization in the $+x$-direction, $\chi_e  = \frac{1}{\sqrt{2}} \left(\begin{array}{c} 1  \\ 1 \end{array} \right)$. 

\begin{figure}
	\includegraphics[scale=0.37,angle=0]{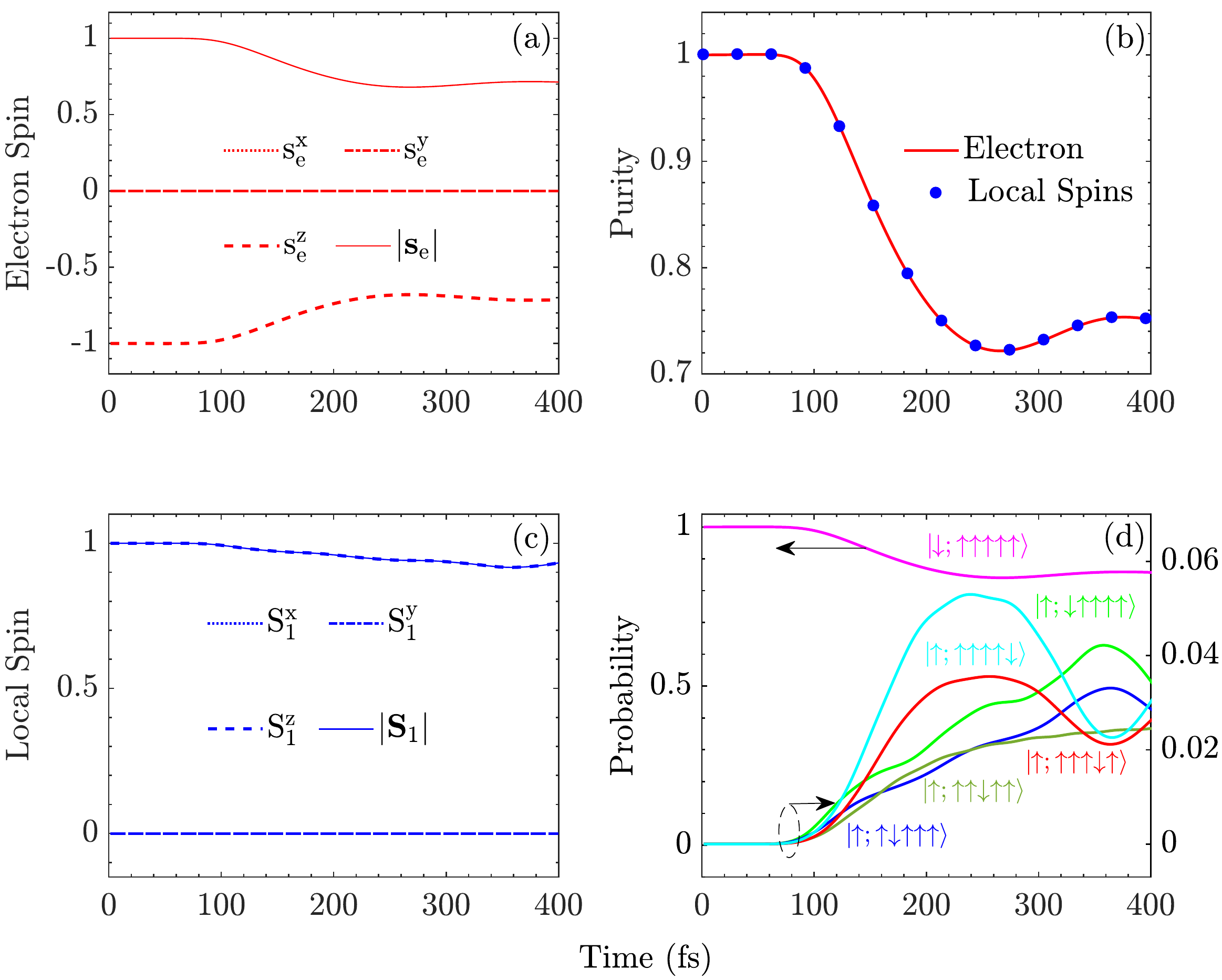}
	\caption{Time dependence of the expectation value of spin (in units $\hbar/2$) obtained from spin-$\frac{1}{2}$ density matrix in Eq.~\eqref{eq:rhospin} for: (a) spin of injected electron wave packet in Fig.~\ref{fig:fig1} which at $t=0$ points in the $-z$-direction that is {\em collinear} and antiparallel to local spins pointing in the $+z$-direction; and (c) first local spin in Fig.~\ref{fig:fig1} [time dependence of expectation value of local spins $n=$2--5 is nearly identical to (c)]. (b) Purity defined in Eq.~\eqref{eq:purity} of the subsystem composed of electron degrees of freedom (orbital and spin) or of the subsystem composed of all local spins. (d) Probability in Eq.~\eqref{eq:prob} to find electron-spin+local-spins subsystem in many-body quantum state $|\sigma_e;\sigma_1\sigma_2\sigma_3\sigma_4\sigma_5\rangle$ .}
	\label{fig:fig3}
\end{figure}
\begin{figure}
	\includegraphics[scale=0.37,angle=0]{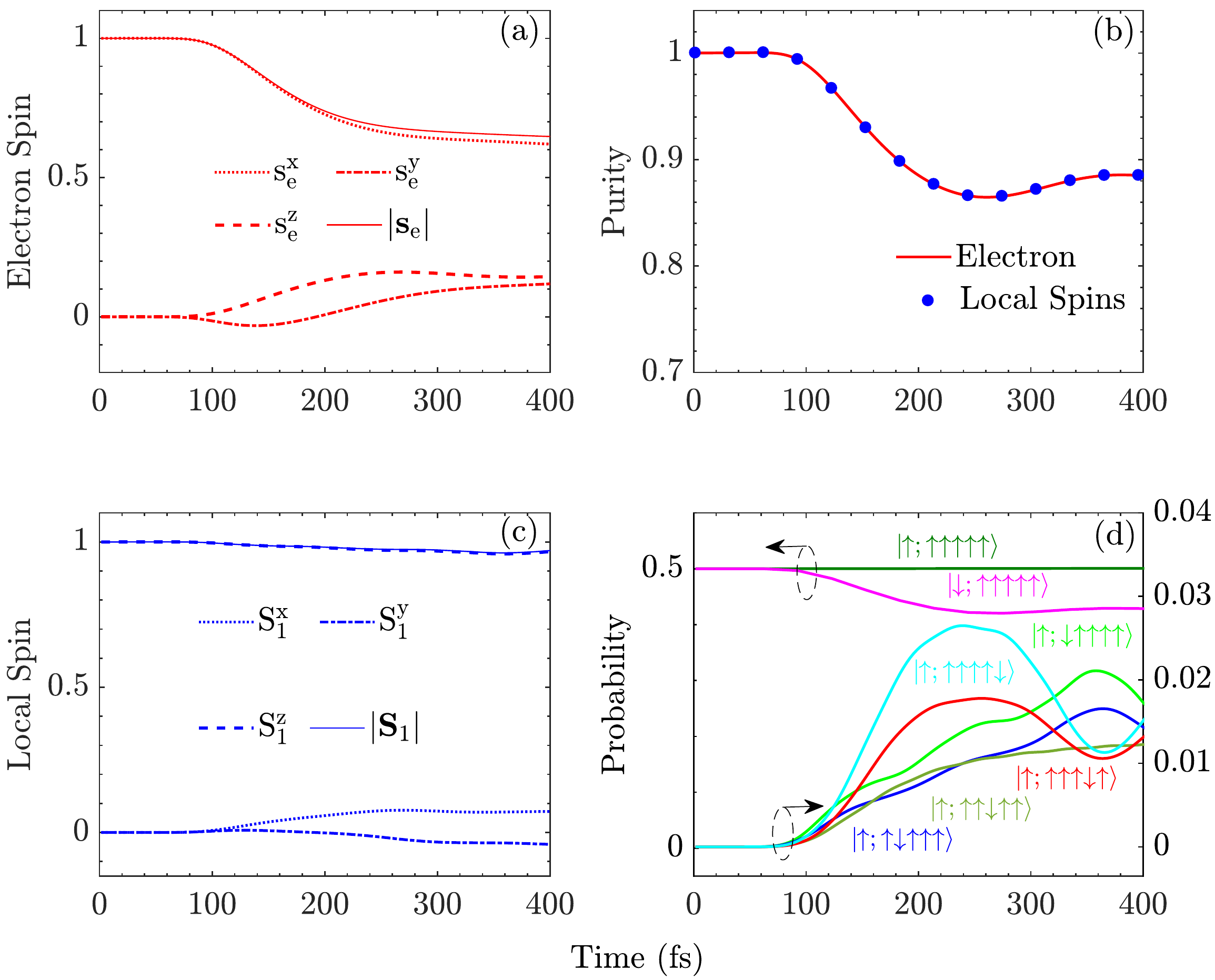}
	\caption{Panels (a)--(d) plot the same information as panels (a)--(d), respectively, in Fig.~\ref{fig:fig3} but for injected electron wave packet which at $t=0$ is spin-polarized in  the $+x$-direction, i.e., {\em noncollinear} to local spins pointing in the $+z$-direction.}
	\label{fig:fig4}
\end{figure}

Although many-body quantum system depicted in Fig.~\ref{fig:fig1} could be evolved via time-dependent density matrix renormalization group~\cite{Schollwoeck2005} for large number of local spins $\sim 100$, for transparency of our discussion, operating with small number of excited states of the XXZ chain which can be analyzed one by one, we employ $N=5$ local spins. The chosen length $L_x=400$ of the TB chain ensures that wave packet does not reflect from its boundaries within the time frame considered in Figs.~\ref{fig:fig3} and ~\ref{fig:fig4}. The {\em numerically exact} $|\Psi(t)\rangle$, governed by the Schrödinger equation $i\hbar \partial |\Psi(t)\rangle/\partial t = \hat{H} |\Psi(t)\rangle$, is obtained by using the Crank-Nicolson algorithm~\cite{Bayfield1999}. 

Figure~\ref{fig:fig2}(c) shows that degenerate ground state has electron and local spins parallel to each other due to $s$-$d$ interaction between them acting to align them. Thus, when an electron with spin-$\downarrow$ along the $-z$-direction is injected, its spin is collinear to local spins but \mbox{$|\Psi(t=0) \rangle = |G \rangle \otimes |\!\!\downarrow_e; \uparrow_1 \cdots \uparrow_N \rangle$} (this form is used below for economy of notation) at $t=0$ is not an eigenstate of the Hamiltonian in Eq.~\eqref{eq:hamil}. {\em This causes time evolution of the electron-spin+local-spins subsystem, which rigorously defines quantum STT even in situation where standard STT in Eq.~\eqref{eq:stt} is identically zero}. In the course of time evolution, $|\Psi(t)\rangle$ becomes an entangled state  due to linear superpositions of separable states being generated for $t>0$. The entanglement entails that each subsystem must be  described using the appropriate reduced density matrix~\cite{Ballentine2014}
\begin{equation}\label{eq:partialtrace}
\hat{\rho}_\mathrm{sub} = \mathrm{Tr}_\mathrm{other} |\Psi(t)\rangle \langle \Psi(t)|, 
\end{equation} 
obtained via partial trace applied to the pure state density matrix $|\Psi(t)\rangle \langle \Psi(t)|$. For example, tracing over the states in the subspace  \mbox{$\mathcal{H}^e_\mathrm{orb} \otimes \mathcal{H}^e_\mathrm{spin} \mathcal{H}_\mathrm{spin}^2 \cdots \otimes \mathcal{H}_\mathrm{spin}^N$}  yields the density matrix of first local spin
\begin{equation}\label{eq:rhospin}
\hat{\rho}_1(t) = \frac{1}{2} \left[\hat{I} + \mathbf{S}_1(t) \cdot \hat{\bm \sigma} \right],
\end{equation}
where $\mathbf{S}_1(t)=\mathrm{Tr}[\hat{\rho}_1(t)\hat{{\bm \sigma}}]$ is the spin expectation value (in units of $\hbar/2$), also denoted as the polarization (or Bloch) vector~\cite{Ballentine2014}. Pure (or fully coherent) quantum states of spin-$\frac{1}{2}$ are characterized by $|\mathbf{S}_1|=1$, while $0<|\mathbf{S}_1|<1$ signifies their decoherence~\cite{Joos2003,Ballentine2014} toward mixed (or partially coherent~\cite{Nikolic2005}) states. Figure~\ref{fig:fig3}(c) shows that first local spin has $S_1^z < 1$, $S_1^x=S_1^y \equiv 0$ and, therefore, $|\mathbf{S}_1|<1$.  The electron spin also exhibits decoherence, $|\mathbf{s}_e|<1$, in Fig.~\ref{fig:fig3}(a). Virtually the same time-dependences as in Fig.~\ref{fig:fig3}(c) are  obtained for other local spins $i=2,\ldots 5$, and, therefore, for total magnetization as the sum of local spins. Thus, this could be precisely the highly nonclassical state of magnetization conjectured from the measurement of the spin valve resistance~\cite{Zholud2017}, which increases $\propto 1 - M_z$ due to magnetization $M_z = g\mu_B\sum_i S_i^z$ shrinking without rotation (i.e., $M_x=M_y=0$) away from its initial orientation.

To explain the origin of magnetization decoherence, or, equivalently, of the subsystem comprised of {\em all} local spins, we view multipartite [due to $N+2$ factors in Eq.~\eqref{eq:hilbert}] total system as a bipartite one, i.e., as being composed of the electron subsystem with states residing in $\mathcal{H}^e_\mathrm{orb} \otimes \mathcal{H}^e_\mathrm{spin}$ and the subsystem of local spins. The purity of the former is defined as~\cite{Ballentine2014,Joos2003}
\begin{equation}\label{eq:purity}
\mathcal{P}^\mathrm{local}_\mathrm{spins}(t)=\mathrm{Tr}\, \left\{ [\hat{\rho}^\mathrm{local}_\mathrm{spins}(t)]^2 \right\}, 
\end{equation}
where density matrix $\hat{\rho}^\mathrm{local}_\mathrm{spins}(t)$ is obtained via Eq.~\eqref{eq:partialtrace} by tracing over the states in the subspace \mbox{$\mathcal{H}^e_\mathrm{orb} \otimes \mathcal{H}^e_\mathrm{spin}$}. The decay of $\mathcal{P}^\mathrm{local}_\mathrm{spins}(t)$ below one in Fig.~\ref{fig:fig3}(b) quantifies ``true decoherence''~\cite{Joos2003} of initially pure state \mbox{$|\!\! \uparrow \rangle \otimes |\!\! \uparrow \rangle \otimes \cdots \otimes |\!\! \uparrow\rangle$} as the decay~\cite{Ballentine2014,Joos2003} of the off-diagonal elements of $\hat{\rho}^\mathrm{local}_\mathrm{spins}(t)$ caused by entanglement with the electron subsystem. The purity of decohered electron subsystem in Fig.~\ref{fig:fig3}(b)  is identical to that of the local spin subsystem, as expected for entanglement in bipartite quantum systems~\cite{Ballentine2014,Joos2003}.

To understand the states of electron-spin+local-spins subsystem which are excited during time evolution initiated by injection of a single spin-polarized electron, we compute the density matrix  $\hat{\rho}_\mathrm{spins}^{\mathrm{e+local}}(t)$ of this subsystem obtained by partial trace in Eq.~\eqref{eq:partialtrace} performed over the states in $\mathcal{H}^e_\mathrm{orb}$. The probability to find this subsystem  in state $|\sigma_e;\sigma_1\ldots\sigma_N\rangle$ at time $t$ 
\begin{equation}\label{eq:prob}
\mathrm{prob}^{\mathrm{e+local}}_\mathrm{spins}(t)=\langle \sigma_e;\sigma_1\ldots\sigma_N| \hat{\rho}_\mathrm{spins}^{\mathrm{e+local}}(t) |\sigma_e;\sigma_1\ldots\sigma_N \rangle,
\end{equation}
is shown in Fig.~\ref{fig:fig3}(d) for electron injected with spin along the $-z$-direction. The subspace of $\mathcal{H}$ whose states can generate nonzero $\mathrm{prob}^{\mathrm{e+local}}_\mathrm{spins}(t)$ is restricted by energy bands in Fig.~\ref{fig:fig2}(b) [caused by anisotropy and boundaries~\cite{Joel2013}] and symmetries, such as that total spin in the $z$-direction has to be conserved in time evolution due to operator $\hat{S}^z_\mathrm{tot}=\hat{s}^z + \hat{S}_1^z + \ldots \hat{S}_N^z$ commuting with the Hamiltonian in Eq.~\eqref{eq:hamil}, $[\hat{H},\hat{S}^z_\mathrm{tot}]=0$. Because of the latter requirement, all states $|\sigma_e;\sigma_1\ldots\sigma_N \rangle$ participating in time evolution must have the same number of $\uparrow$-spins, so that one finds in Fig.~\ref{fig:fig3}(d) progressive excitation of states with flipped spin of electron and one flipped local spin with transfer of angular momentum of $1 \times \hbar$. However, the initial state $|\downarrow; \uparrow \ldots \uparrow \rangle$ maintains its probability close to one, and other states with flipped electron spin and one flipped local spin have much smaller and nonuniform probability. Such peculiar quantum superposition of separable many-body states, with large contribution from the initial state,  leads to local spins maintaining their direction along the $z$-axis in Fig.~\ref{fig:fig3}(c). This can be contrasted with na\"{i}ve (i.e., not taking into account superpositions) intuition~\cite{Gauyacq2011,Balashov2008} where spin-$\downarrow$ electron simply flips first local spin---the flip then propagates to displace transversally other local spins away from the anisotropy axis, eventually exciting white spectrum~\cite{Gauyacq2011} of lowest spin waves modes~\cite{Parkinson2010} (where total spin 
is lower than that of the ground state by $1 \times \hbar$) of the free FM layer.  

The same effects---entanglement of electron state and state of all local spins [Fig.~\ref{fig:fig4}(b)]; thereby induced decoherence of electron spin [Fig.~\ref{fig:fig4}(a)] and local spins [Fig.~\ref{fig:fig4}(c)]; and high probability [Fig.~\ref{fig:fig4}(d)] to find initial state of electron-spin+local-spins subsystem in the course of time evolution---is present also in standard STT geometry with noncollinearity between spin of injected electron and local spins. Furthermore, probabilities  $\mathrm{prob}^{\mathrm{e+local}}_\mathrm{spins}(t)$ in Fig.~\ref{fig:fig4}(d) to excite states of type \mbox{$|\!\! \uparrow; \uparrow \ldots \downarrow \ldots \uparrow \rangle$} are simply half of those obtained for collinear geometry in  Fig.~\ref{fig:fig3}(d)  since spin of injected electron along the $+x$-direction used in Fig.~\ref{fig:fig4}  means \mbox{$|\!\! \rightarrow_e \rangle=\frac{1}{\sqrt 2}(|\!\!\uparrow_e\rangle + |\!\!\downarrow_e \rangle)$} where only $\frac{1}{\sqrt 2}|\!\!\downarrow_e \rangle$ 
term, entering as a factor of the initial many-body state \mbox{$|G\rangle \otimes \frac{1}{\sqrt 2}(|\!\!\uparrow_e\rangle + |\!\!\downarrow_e \rangle) \otimes |\!\! \uparrow_1 \ldots  \uparrow_N \rangle$}, induces time evolution of local spins and transfer of angular momentum. On the other hand, \mbox{$\frac{1}{\sqrt 2}|\!\!\uparrow_e\rangle \otimes |\!\! \uparrow_1 \ldots  \uparrow_N \rangle$} term in the initial many-body state is an eigenstate [Fig.~\ref{fig:fig2}(c)] of electron-spin+local-spins subsystem and, therefore, has $\mathrm{prob}^{\mathrm{e+local}}_\mathrm{spins}(t) = 1/2$ which does not evolve in time in Fig.~\ref{fig:fig4}(d). Thus, identical profile of curves in Figs.~\ref{fig:fig3}(d) and ~\ref{fig:fig4}(d) reveal that {\em in fully quantum many-body picture there is no difference between standard STT and quantum STT}---both require $|\!\!\downarrow_e \rangle$ factor state in the  initial many-body state, where such factor is due to electron spin state (in the collinear case) or a term in the superposition of electron spin states (in the noncollinear case).

{\em Note added.}---During the completion of this work, we became aware of two studies~\cite{Qaiumzadeh2018,Bender2018} where magnetization dynamics in collinear spin valves at cryogenic temperatures is attributed to current-enhanced quantum spin fluctuations. However, as discussed above, such fluctuations are forbidden in the ground state of ferromagnets~\cite{Yosida1996}, and  if generated as random magnetic field~\cite{Chudnovskiy2008} by the nonequilibrium spin shot noise~\cite{Qaiumzadeh2018}, they would lead to rotation of magnetization away from the initial collinear direction that was excluded from the experiments of Ref.~\cite{Zholud2017}. We propose to clarify the role of spin shot noise by using a sequence of  Lorentzian voltage pulses   to inject a train of levitons into a collinear spin valve, where leviton as a minimal collective many-body excitation above the Fermi sea carrying a single electron charge, is free of particle-hole pairs and has vanishing current noise~\cite{Keeling2006,Dubois2013}.

\begin{acknowledgments}
We are grateful to S. Urazhdin for instructive discussions. P. M. and B. K. N. were supported by NSF Grant No. ECCS 150909. M.~P. and P.~P. were supported by ARO MURI Award No. W911NF-14-0247. 
\end{acknowledgments}




\end{document}